\documentclass[smallextended]{svjour3}       
\smartqed  
\usepackage{latexsym}

\usepackage{graphicx}

\newcommand{\be}{\begin{equation}}

\newcommand{\ee}{\end{equation}}

\newcommand{\reff}[1]{(\ref{#1})}

\begin{document}

\title{From the pion cloud of Tomonaga to the electron pairs of Schrieffer: many body wave functions from nuclear physics to condensed matter physics}


\titlerunning{From Tomonaga nucleon pion cloud theory to BCS theory and beyond}       

\author{Fabrizio Palumbo  \and Augusto Marcelli \and Antonio Bianconi}

\institute{Fabrizio Palumbo  \at INFN Laboratori Nazionali di Frascati, c.p. 13, I-00044 Frascati, Italy \\ \email{fabrizio.palunbo@lnf.infn.it} \and
         Augusto Marcelli \at INFN Laboratori Nazionali di Frascati, c.p. 13, I-00044 Frascati, Italy $and$ RICMASS, Rome International Center for Materials Science Superstripes, 00185 Rome, Italy  \\ \email{augusto.marcelli@lnf.infn.it}  \and Antonio Bianconi \at RICMASS, Rome International Center for Materials Science Superstripes, Via dei Sabelli 119A, 00185 Rome Italy \\ Istituto di Cristallografia, CNR, Montelibretti, Rome, Italy \\ 
         National Research Nuclear University MEPhI, Kashirskoe shosse 31 Moscow, 115409, Russia  \\ \email{antonio.bianconi@ricmass.eu} }
         
 \date{Received: date / Accepted: date}

\maketitle

\begin{abstract}

It is well known that diverse pieces of models and physical ideas coming from different areas of physics converged in the BCS theory of superconductivity. On the contrary it is little known that the formalism developed in the Tomonaga quantum field theory of the pion-nucleon system was an important ingredient for the development of BCS theory. We discuss the evolution of these ideas in quantum field theory providing an unconventional historical perspective.



\end{abstract}

\section{Introduction}

Diverse pieces of models and physical ideas coming from different areas of physics converged into the BCS theory of superconductivity \cite{BCS,BCS1,Schr1}. The previous contributions of F. London \cite{London} for our understanding of the superconducting state and the electron-phonon theory of H. Fröhlich \cite{Frohlich} are well known. The theory of Bose-Einstein condensation \cite{Bose,Einstein}, the two-fluid model of Tisza \cite{Tisz} and the Landau \cite{Land} theory of superfluidity clearly paved the way to the BCS theory. However, as far as we understand, the two-fluid model did not play any role in its construction and the BCS authors claimed not to have used ideas of Bose-Einstein condensation and superfluidity. Actually their relevance to the theory of superconductivity was a matter of debate with the Sidney group of Schafroth and Blatt \cite{Blat}. 
In a Physics Today paper published in 1992 \cite{Schr} Schrieffer mentions the "intense" discussions they had with Blatt before their theory was formulated in connection with the size of the Cooper pairs and the nature of the ground state in superconductors. He reports that Blatt thought of electron pairs of small size, while Bardeen maintained that they should be large. It seems that the underlying idea was that only small pairs might undergo a Bose-Einstein condensation (BEC). In such a case the quotation of the discussions with Blatt might have been an indirect answer to the strongly polemic comment that Blatt made in his beautiful book on superconductivity \cite{Blat}, stating that BCS did not really understand or at least did not appreciate that their theory actually involves Bose-Einstein condensation of Cooper pairs. We emphasize that even though the quoted paper by Schrieffer follows by 28 years the Blatt book, Schrieffer does not mention it explicitly in his report about their diverging opinions. 

Much less known is the role of the advances in meson theory of nuclear physics \cite{Wentzel} and in particular of the Tomonaga model of the nucleon-pion system \cite{Tomo} even though Schrieffer acknowledged this work in the paper written in memory of Bardeen \cite{Schr}. 
Also the development of field theoretical methods in meson physics \cite{Wentzel} and in particular the Tomonaga model of the nucleon-pion system \cite{Tomo}, 
were addressed to the study of the strong \cite{Wentzel} and intermediate \cite{Tomo} interactions. 
The formalism of the nucleon theory of Tomonaga in the $intermediate$ coupling regime was indeed an essential ingredient for the construction of the BCS wave function \cite{Schr} in the weak coupling regime. 

Today in the field of high temperature superconductivity \cite{Bia1,Bia2,Bia3,Buss,Gui} and in the quantum condensation of fermionic ultracold gases \cite{Perali,Par,Sha,Leg} there is high interest in superconductivity in strongly interacting fermionic systems where quantum condensation takes place in a regime of $intermediate$ coupling between the BCS and the BEC regime, called BCS-BEC crossover.

It is our purpose to give some details about the above issues, essentially based on the quoted paper by Schrieffer and the book by Blatt. We do not make any attempt at an accurate and thorough historical research to fully settle this matter. We discuss some aspects of the mentioned facts from a historical perspective.

\section{Tomonaga and BCS wave functions}

According to Schrieffer's recollection \cite{Schr} about how the theory of superconductivity was constructed,  the breakthrough
in devising the correct many-electrons wave function came in a way that appears almost fortuitous.
Schrieffer reports that attending a meeting on the many-body theory in 1957 it occurred to him \cite{Schr} that  \\ 
"  because of the strong overlap of pairs perhaps a statistical approximation analogous to a type of mean field would be appropriate to the problem. Thinking back to a paper by Sin-Itiro Tomonaga that described the pion cloud around a static nucleon \cite{Tomo}" \\ 
he tried a similar wave function for the electron system.

Tomonaga investigated \cite{Tomo} the structure of the wave function of a nucleon at rest interacting with pions (called then mesotrons). The state of a physical nucleon is constructed in terms of bare nucleons surrounded by virtual pions. Neglecting for simplicity neutral pions, the wave function of a physical proton, for instance,  has two components: one representing a bare proton with a cloud of an equal number of positive and negative pions, the other one representing a bare neutron accompanied by a positive pion plus a cloud of an equal number of positive and negative pions
\begin{eqnarray}
\Psi_1 &=& \sum_n \phi_n(k^+_1,... k^+_n; k^-_1,...k^-_n) 
\nonumber\\
\Psi_2&=&\sum_n \psi_n(k^+_1,... k^+_{n+1} ; k^-_1,...k^-_n) 
\label{T}
\end{eqnarray}
 where $k^{\pm}$ are the momenta of the positive/negative pions. The structure of the pions cloud was studied in the Hartree approximation
 \begin{eqnarray}
&& \phi_n(k^+_1... k^+_n; k^-_1...k^-_n) = \mbox{const}
 \nonumber\\
 &&\,\,\,\,\,\,\, \,\,\,\,\,\,\times f_+(k_1^+)... f_+(k_n^+)f_-(k_1^-)... f_-(k_n^-)
\nonumber\\ 
 && \psi_n(k^+_1... k^+_{n+1}; k^-_1...k^-_n) = \mbox{const}
 \nonumber\\
 &&\,\,\,\,\,\,\, \,\,\,\,\,\,\times f_+(k_1^+)... f_+( k^+_{n+1})f_-(k_1^-)... f_-(k_n^-)\label{Hartree1}
\end{eqnarray}
choosing a parametric form for the functions $f_{\pm}$.

The BCS theory focussed on some elements that a microscopic theory should contain, among which of the highest consequence the fact that the wave function should have all the matrix elements with the relevant, attractive part of electron-electron interaction potential, of the same sign. 
Such a wave function should automatically produce a gap in the quasiparticle spectrum. 

This can be achieved by occupying the single-electron states in pairs. The pair states should be chosen in such a way that transitions between them can be induced by a translation invariant potential, and therefore all the pairs should have one and the same total momentum. To form the ground state the best choice is to associate the states ${\bf k} \uparrow$ with the states $- {\bf k} \downarrow$ since the exchange terms reduce the matrix elements between states with parallel spins. It is impressive how these authors arrived at contriving so tightly the theory starting from phenomenology. 

BCS started from the general form \cite{BCS} 
\be
\Psi = \sum_n b_n(\bf{k}_1,... k_n)\label{BCS}
\ee
where $\bf{k}_i$ represents a pair of electrons with opposite spins and opposite momenta equal to $\pm {\bf k}_i$. For the calculations a Hartree-like approximation was adopted, setting
\be
b_n({\bf k}_1,... {\bf k}_n)= b({\bf k}_1)... b({\bf k}_n). \label{Hartree2}
\ee
This is not exactly a Hartree approximation, because $b(\bf{k}) $ does not describe a single particle, but an electron pair respecting the Pauli principle. Nevertheless the simplicity of an independent particle calculation is preserved, because non-vanishing matrix elements of the two-body potential connect only configurations that differ by only one of the occupied pairs.

We can understand how the wave function~\reff{Hartree1}  might have inspired Schrieffer to write the wave function~\reff{Hartree2}, that can be obtained  by setting in~\reff{Hartree1}
\begin{eqnarray}
b({\bf k}_i) &=& f_+({\bf k}_i^+) f_-({\bf k}_i^-)
\nonumber\\
{\bf k}^-_i &=&-{\bf k}^+_i = - {\bf k}_i
\end{eqnarray}
with the understanding that the $f_{\pm}$ describe here electrons with $up/down$ spins while in \reff{Hartree1} they describe pions with $up/down$ isotopic spin.

It is perhaps worth while noticing that the use of the first quantization in both cases makes these wave functions more similar. 

The BCS paper that followed their Letter only few months later \cite{BCS1}, however, was formulated in second quantization
\be
| \Psi >= 1 + \sum_{K>0} \frac{u_K}{v_K} \, c_K^{\dagger} c_{-K}^{\dagger} +
\sum_{K>0} \frac{u_K}{v_K} \frac{u_L}{v_L}\, c_K^{\dagger} c_{-K}^{\dagger} c_L^{\dagger} c_{-L}^{\dagger}|0>\,.
\ee
In this form the similarity between the Tomonaga and the BCS wave functions looks less impressive, but once the above wave function is rewritten in the form
\be
| \psi >= \prod_{K>0} (u_K + v_K \, c_K^{\dagger} c_{-K}^{\dagger}) |0>
\ee
it is easy \cite{Blat} to recognize that its component with $N$ electrons
\be
| \Psi_N >= \left( \sum_K\frac{u_K}{v_K} \, c_K^{\dagger} c_{-K}^{\dagger}\right)^\frac{N}{2} |0>
\ee
describes a condensate of electron pairs in the quantum state
\be
|\phi> = \sum_K\frac{u_K}{v_K} \, c_K^{\dagger} c_{-K}^{\dagger} |0> \label{projected}
\ee
 
even though the total wave function represents pairing of single electron states (but Blatt writes: even if its projected component represents pairing of single electron states, rather than a quantum state of an electron pair, Blatt, p. 175).

 This property was recognized independently by F. J. Dyson, M. R. Schafroth and B. F. Bayman, see footnote 7, p. 182 in the Blatt book, showing that the Bose-Einstein condensation is indeed a feature present in the BCS theory. 

The momenta of the electrons of the condensate are contained in a tiny shell around the Fermi momentum. Therefore the BCS wave function describes a two fluid system: the normal fluid made of the unpaired electrons and the superfluid made of the condensed pairs.

\section{Bose-Einstein condensation and the theory of quasi-chemical equilibrium}

The idea that condensation of some sort was at the basis of the mechanism leading
to superconductivity appeared independently at various stages in the community investigating
 this subject. It already appeared in a paper by London \cite{Lond} in 1938, and it was first formulated in a clear way in 1946 by Ogg, even in the title 
of his paper: "Bose-Einstein condensation of trapped
 electron pairs" \cite{Ogg}. Ogg claimed the observation of persistent electric
 currents in very dilute solutions of alkali metals in liquid ammonia at temperatures of the order of 180 C.
 He explained such a phenomenon in terms of trapped electron pairs. 
 Some attempts to reproduce his results, however, gave contradictory results \cite{Boor,Ho,Gi}, and apparently 
 the whole thing did not receive any further attention and therefore did not play any role in the 
 development of the microscopic theory of superfluidity, apart from a possible subconscious
 influence on the Sidney group, as reported by Blatt \cite{Blat}. It seems that what remained of this research is the Gamow's limerick reported by P. T. Landsberg in his correspondence with Blatt \cite{Blat}

  In Ogg's theory it was his intent
  
  that the current keep flowing, once sent;
  
  so to save himself trouble,
  
  he put them in double,
  
  and instead of stopping, it went.

The fact that this matter has not been settled up to now, if true, is surprising to us, the more so since Ogg's result is sometimes regarded as the first evidence of strong coupling condensates found in high temperature superconductors. \cite{Bia1,Bia2,Cai,Phi,Bia4}.

Condensation of bound electron pairs was related to superconductivity also by Schafroth \cite{Scha}, whose work did instead certainly play a major role in the Sidney group activity. Actually the Schafroth papers went much further, pointing out that superconductivity 
occurs in gas of charged bosons in the presence of Bose-Einstein condensation and suggesting that bound states of electron pairs should 
condense in superconductors. It is remarkable that also Ginsburg "as early as 1952 noted that the charged Bose gas would behave as a superconductor, but did not arrive at the idea of pairing" \cite{Gins}.

Pair condensation was at the basis of the so called quasi-chemical equilibrium theory formulated by the Sidney group. According to this theory in superconductors there is a dynamical equilibrium between a gas of electrons and a gas of bound electron pairs, in analogy with the phenomenological two-fluid model. The ground state wave function of the quasi-chemical equilibrium theory reads
\be
\Phi={\mathcal A}\{\phi(1,2) \phi(3,4)...\phi(N-1,N)\} \label{Blatt}
\ee
where ${\mathcal A} $ is the antisymmetrizer operator and $\phi(i,j)$ is the wave function of electrons $i,j$. At the end of the calculations it was recognized that this wave function is equal to the projected component of the BCS wave function of Eq.\reff{projected} and this theory produced exactly the same results as the BCS one. 

 At this point an important comment is in order. The wave functions~\reff{T} and ~\reff{BCS} are superpositions of states with different number of particles. There is
nothing against this when the particles are mesons, but this is forbidden in the case of electrons by a superselection rule. The formalism of the quasichemical equilibrium theory instead respects exactly fermion number conservation. At the time this might have appeared an advantage, but not only fermion number conservation did make the calculations much more complicated, in addition its violation opened the way to the discovery of spontaneous symmetry breaking that was of the highest consequence in the development of quantum field theory and condensed matter physics.

 Also the BCS theory uses the world "condensation" but in a way less clear to us: they think of a condensation of some kind, "condensation of the average momentum distribution", "condensation in momentum space" [\cite{Schr}, but keep away from bosons, for which there is no gap in the quasiparticles spectrum. 
According to Schrieffer BCS "were aware of the efforts by Blatt and Schafroth, who were working with the quasi-chemical theory " that according to Schrieffer's recollection " corresponded to" the limit of a very weakly overlapping of electron pairs, strongly bound, which Bardeen thought was physically untenable, and they had intense discussions on this point with Blatt during his visit at Urbana {\cite{Schr}". Their main objection to Bose-Einstein condensation was that the electron pairs should be large and therefore largely overlapping \cite{Schr}, rather than a gas of tightly bound states, but we did not find a place where the Sidney group advocates that the size of the electron pairs should be small. Instead Blatt \cite{Blat} on p. 182 reports a sentence by Cooper \cite{Coop} "...the deep mystery in the theory of superconductivity, the pairing condition" with the comment :" It is unusual to find, five years after the original publication of a theory, an admission by one of the authors that the basic assumption of the theory is a deep mystery to him".

The emphasis on the relevance of Bose-Einstein condensation and the relative controversy goes further in connection with the explanation of persistent currents. For Blatt "Bose-Einstein-like condensation into a state of macroscopic de Broglie wave length " provides the qualitative explanation of superconductivity, because then the scattering centers (impurities, lattice defects...) affect the supercurent in a coherent fashion.
The noncoherent scattering is proportional to the fluctuation of the number of scattering centers in a volume equal to $\lambda^3$, where $\lambda$ is the relevant wavelength that is equal to the linear dimension of the specimen, so that this fluctuation vanishes making the lifetime of the supercurrent infinite. The existence of a gap is irrelevant to such a mechanism. Therefore Blatt concludes (Blatt, p. 318) that the theory of persistent currents developed by Bardeen \cite{Bard} based on the existence of a gap is "basically incorrect " depending on an "invalid application of pertubation theory".

In any case the size of electron pairs is not an intrinsic feature of the quasichemical equilibrium theory, it depends on the parameters of the theory. Moreover, if we define "condensation" a state in which there is a macroscopic occupation of one and the same single particle state, we should conclude that the BCS state does describe a condensate. It is remarkable in this connection that in 1958 Bogoliubov and collaborators 
published a paper \cite{Bogo} in which they explain the persistent currents in the same way as 10 years before Bogoliubov explained the frictionless flow in a superfluid: the energy excitations in both the supercurent and the frictionless flow states are positive below some critical velocity, making it impossible for such states transitions that would deplete them.

 If the view that both BCS and quasichemical equilibrium theories describe a Bose-Einstein condensate is accepted, what is called BCS-Bose-Einstein Condensation crossover 
should be called " fermion pairs size crossover ", the size being large in the BCS state and small in the Bose-Einstein condensate. These states can be described by wave functions of the same structure. Indeed Legget \cite{Legg} advocates the convenience of the quasichemical equilibrium theory to describe such crossover.

In conclusion the forgotten paper by Tomonaga is a key missing piece to understand the evolution of the quantum field theory of quantum condensation in the $intermediate$ coupling regime which could help to better understand the BCS-BEC crossover regime also in the pressurized sulfur hydrides with T$_c$=203K \cite{H3S1,H3S2,H3S3}.

\end{document}